# Guiding Symbolic Execution with Static Analysis and LLMs for Vulnerability Discovery

Md Shafiuzzaman
University of California, Santa Barbara
Santa Barbara, CA, USA
mdshafiuzzaman@ucsb.edu

Achintya Desai
University of California, Santa Barbara
Santa Barbara, CA, USA
achintya@ucsb.edu

Wenbo Guo
University of California, Santa Barbara
Santa Barbara, CA, USA
henrygwb@ucsb.edu

Tevfik Bultan
University of California, Santa Barbara
Santa Barbara, CA, USA
bultan@ucsb.edu

## Abstract

Symbolic execution detects vulnerabilities with precision, but applying it to large codebases requires harnesses that set up symbolic state, model dependencies, and specify assertions. Writing these harnesses has traditionally been a manual process requiring expert knowledge, which significantly limits the scalability of the technique. We present Static Analysis Informed and LLM-Orchestrated Symbolic Execution (Sailor), which automates symbolic execution harness construction by combining static analysis with LLM-based synthesis. Sailor operates in three phases: (1) static analysis identifies candidate vulnerable locations and generates vulnerability specifications; (2) an LLM uses vulnerability specifications and orchestrates harness synthesis by iteratively refining drivers, stubs, and assertions against compiler and symbolic execution feedback; symbolic execution then detects vulnerabilities using the generated harness, and (3) concrete replay validates the symbolic execution results against the unmodified project source. This design combines the scalability of static analysis, the code reasoning of LLMs, the path precision of symbolic execution, and the ground truth produced by concrete execution. We evaluate Sailor on 10 open-source C/C++ projects totaling 6.8 M lines of code. Sailor discovers 379 distinct, previously unknown memory-safety vulnerabilities (421 confirmed crashes). The strongest of five baselines we compare Sailor to (agentic vulnerability detection using Claude Code with full codebase access and unlimited interaction), finds only 12 vulnerabilities. Each phase of Sailor is critical: Without static analysis targeting confirmed vulnerabilities drop 12.2×; without iterative LLM synthesis zero vulnerabilities are confirmed; and without symbolic execution no approach can detect more than 12 vulnerabilities.

## 1 Introduction

Automated vulnerability detection in large C/C++ codebases is a long-standing challenge. Modern static analysis (SA) tools [1, 3, 4, 8] can scan millions of lines and efficiently flag thousands of candidate sites, but with high false positive rates. Fuzzing tools [11, 26] complement SA by exercising the program with concrete inputs, yet they struggle to reach deep library internals that require precisely structured program state. Recently, large language models (LLMs) have been applied to vulnerability detection [10, 17, 30? ], but their outputs carry no formal execution-based verdict, a model may confidently report vulnerabilities that do not exist or dismiss genuine vulnerabilities as safe.

Symbolic execution (SE) [7, 18] offers certain advantages compared to these techniques: It treats inputs as symbolic values, it can target functions directly without requiring end-to-end program execution, and it uses a constraint solver to produce concrete *witness inputs* that formally confirm or refute a property violation. However, SE cannot be applied to most codebases directly. It faces significant scalability challenges: Path explosion limits the depth of exploration, environment dependencies (system calls, file I/O, complex library APIs) require modeling or abstraction, and the SE engine cannot determine *where* in a large codebase to begin searching. In practice, these challenges are addressed through a well-formed *harness*, a driver program that selects an entrypoint and sets up symbolic inputs (targeting), abstracts irrelevant code paths and environment dependencies through stubs (path explosion and environment modeling), and instantiates project-specific types with the correct field layout. Constructing such harnesses manually demands thorough knowledge of both the target codebase and the SE engine and constitutes the primary bottleneck that limits the practical adoption of SE at scale. Hybrid approaches that combine SA or fuzzing with SE have been developed [5, 6, 20, 27, 29] to prune search spaces, but the harness construction problem has remained challenging: Either the complete harness (or significant parts of it) is written manually or the system relies on predefined harness templates with manually specified entrypoints and environment model. In this paper, we present a novel approach that addresses the harness generation problem to enable scalable SE for large codebases. We present SAILOR (**S**tatic **A**nalysis **I**nformed and **LL**M-**OR**chestrated Symbolic Execution), a fully automated pipeline that consists of three phases:

(1) *SA informed target generation* (§3): A configurable SA phase (based on an extensible query suite that captures vulnerability patterns) scans the project source and produces vulnerability specifications for potential vulnerabilities.

(2) *LLM-orchestrated SE* (§4): For each vulnerability specification generated in the first phase, an LLM iteratively synthesizes a harness (driver, stubs, and assertions) against compiler and SE feedback. SE engine then uses the harnesses to validate (or eliminate as a false alarm) vulnerability specifications and produces witnesses for validated vulnerabilities.

(3) *Concrete validation* (§5): Witness inputs are replayed against the *unmodified* project source under AddressSanitizer, producing a verdict independent of both the LLM and the harness (in order to eliminate false positives that may be due to unrealistic harness generation by LLM in the second phase).

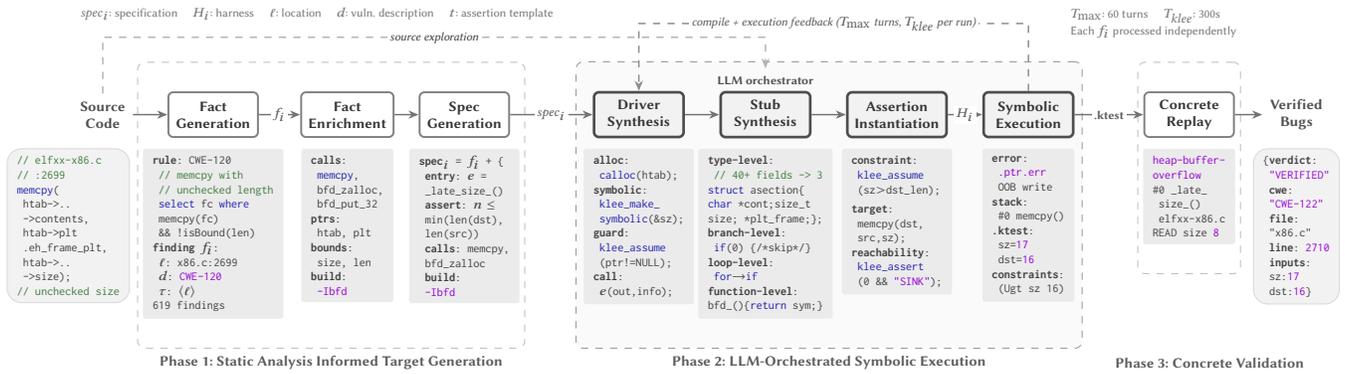

**Figure 1: SAILOR pipeline with running example. Phase 2 (shaded) is the core contribution: the LLM synthesizes and iteratively refines the harness against compiler and KLEE feedback.**

We implemented SAILOR focusing on memory safety using CodeQL [1], KLEE [7], and AddressSanitizer [28], and evaluated it on 10 C/C++ projects (6.8 M LOC). SAILOR discovers **379 unique previously unknown vulnerabilities** (421 confirmed crashes, deduplicated by file, function, and line). This paper makes the following contributions:

- A fully automated vulnerability detection technique that uses SA to transform a project's source code into SE targets, enables an LLM to synthesize SE harnesses through iterative refinement, and confirms each finding via concrete replay against unmodified project binaries.

- SAILOR, an end-to-end implementation of this technique combining SA (CodeQL), an LLM, SE (KLEE), and concrete validation (AddressSanitizer), requiring no manual harness writing or project-specific configuration beyond a build script.

- An evaluation on 10 open-source C/C++ projects demonstrating scalability (63K to 1.84M LOC), precision (379 unique vulnerabilities confirmed by concrete replay), effectiveness (over 30× more confirmed vulnerabilities than the strongest baseline), and reproducibility (each vulnerability is accompanied by path constraints, crashing inputs, ASan crash stack trace, and, where applicable, a fuzz reproduction seed).

## 2 Overview

In Figure 1 we show the SAILOR pipeline using a zero-day vulnerability that SAILOR discovered in GNU Binutils (commit b2bc71a, March 2026) as an example. Binutils is a 1.84 M-line codebase which is not feasible to analyze using manual harness construction. The vulnerability resides in `_bfd_x86_elf_late_size_sections()` in `elfxx-x86.c`, an internal module of the binutils linker. The `memcpy` at line 2699 (Listing 1) copies `size` bytes into a heap-allocated destination `contents`. Two null checks (lines 2696–2697) guard against null-pointer dereferences but do not enforce that `size` is within the bounds of `contents`. When `size` exceeds the allocation, the copy accesses memory beyond the heap boundary (CWE-122, heap-based buffer overflow). Automatically finding such a vulnerability requires answering three questions: **where** does the vulnerability reside in the codebase, **how** the program's behavior leads to the vulnerability, and **whether** that behavior is feasible in a concrete execution. Existing individual vulnerability analysis techniques can only address part of the problem:

```
2696  if (htab->plt_eh_frame != NULL
2697        && htab->plt_eh_frame->contents != NULL)
2698  {
2699      memcpy(htab->plt_eh_frame->contents, // dst
2700          htab->plt_eh_frame_plt,          // src
2701          htab->plt_eh_frame->size);       // size: unchecked
2702  ...
2703
```

**Listing 1: Vulnerable `memcpy` call site in `elfxx-x86.c`.**

*(i) SA* can be used to identify *where* a potential vulnerability exists. For example, a CodeQL unchecked-length pattern flags this `memcpy` as a potential buffer overflow, one of 619 findings from the same rule across binutils. However, it cannot determine whether `plt_eh_frame->size` ever exceeds the memory allocation.

*(ii) Fuzzing* exercises binutils through command-line entrypoints by mutating ELF input files, but the vulnerable function resides deep in the x86-specific linker module, which the OSS-Fuzz integration for binutils[1] explicitly *disables* (`--disable-ld`). Even if enabled, triggering the overflow requires an internal state in which `plt_-eh_frame->size` exceeds the allocation of `contents`: a condition that depends on the linker's internal struct fields derived indirectly from the ELF input.

---

[1] https://github.com/google/oss-fuzz/tree/master/projects/binutils

*(iii) LLM-based detection* may identify the unchecked `memcpy` within a code snippet, and with tool-assisted source exploration can read type context, understand field relationships, and generate code that instantiates complex project-specific structures. However, a 1.84 M-line codebase far exceeds any context window, and confirming the overflow requires determining that `size` can exceed the allocation of `contents` under some feasible execution path. Determining this requires path constraint extraction and constraint solving rather than pattern matching.

*(iv) SE* could derive the overflow condition, but constructing the required harness for this target is non-trivial. The harness must instantiate `elf_x86_link_hash_table`, a deeply nested structure comprising over 40 fields including embedded sub-structs, function pointers, and linker-global state, and must declare `plt_eh_-frame->size` as symbolic with constraints that direct execution toward the overflow path. This demands knowing which fields are relevant, which may be safely zero-initialized, and which can be exercised with concrete values.

No single technique can answer all three questions that are needed to automatically identify the vulnerability. Sᴀɪʟᴏʀ composes multiple techniques in order to resolve all questions: SA resolves the **where** question by searching for vulnerability patterns (Phase 1), the LLM resolves the **how** question by synthesizing project-specific drivers, stubs, and assertions (Phase 2), and SE resolves the **whether** question by deriving path constraints and producing witness inputs (Phase 2), which are then validated by concrete replay (Phase 3). The subsequent sections detail each phase of Sᴀɪʟᴏʀ (§3–§5), using the vulnerabilty highlighted in Figure 1 as a running example.

## 3 Phase 1: Static Analysis Informed Target Generation

Given a project's source code, Phase 1 of Sᴀɪʟᴏʀ generates a set of SE *targets*, each defined by a *vulnerability specification* that identifies a candidate location $\ell$, the SA evidence supporting it, and the contextual information that Phase 2 requires to synthesize a harness. This phase proceeds in three stages: fact generation, fact enrichment, and specification generation (Figure 1).

### 3.1 Fact Generation

Sᴀɪʟᴏʀ builds a CodeQL database from the project's source and build configuration, then runs a suite of 34 memory-safety queries (Table 1): 13 standard queries from the `codeql/cpp-queries` pack and 21 custom queries. Each query produces a SARIF (Static Analysis Results Interchange Format) [25] finding $f_i = (\ell, d, \tau)$ per flagged location:

- *Candidate location* $\ell$: the file, line, and column range identifying the suspected vulnerable site.
- *Vulnerability description* $d$: a natural-language characterization of the vulnerability at $\ell$.

- *Data-flow trace* $\tau = \langle s, \ldots, \ell \rangle$: an ordered sequence of program points from a data-flow source $s$ to the sink at $\ell$. The source $s$ is defined by the query's `isSource` predicate:
  - *Inter-procedural queries* (e.g., use-after-free: $s$ is the `free` call, $\ell$ is the subsequent dereference): $\tau$ contains the full path from $s$ to $\ell$.
  - *Local pattern-based queries* (e.g., buffer overflow via unchecked `memcpy` length): $\tau$ reduces to $\langle \ell \rangle$.

The custom queries address three gaps: (i) missing patterns: for example, the standard CWE-120 queries detect out-of-bounds array accesses such as `buf[i]` where `i` exceeds the array size, but do not flag calls like `memcpy(dst, src, len)` where `len` may exceed the allocated size of `dst`, as in the running example. Our custom rules add this pattern as well as `sprintf`/`snprintf` stack buffer overflows and out-of-range pointer offsets; (ii) finer-grained use-after-free variants (`free`-then-dereference, `free`-then-call-argument, `realloc`-induced stale pointers) that produce more specific descriptions $d$ than the single standard `cpp/use-after-free` query; and (iii) lifetime and type-confusion patterns, including dangling pointers after `free` and use of memory reclaimed by a different type. Since Sᴀɪʟᴏʀ only requires SARIF-formatted findings, the query suite can be extended to other vulnerability classes by adding new queries.

Listing 2 illustrates two custom queries. Part (a) selects calls to memory-copy functions whose length argument is neither guarded by `sizeof` nor bounded by `strlen`; the `select` clause emits the vulnerability description $d$ that Phase 2 passes to the LLM. Applied to binutils, this query flags the `memcpy` at line 2699 in `elfxx-x86.c` (Listing 1), producing $\ell = \texttt{elfxx-x86.c:2699:7}$ and $d = \texttt{"CWE-120:}$ Buffer Overflow via memcpy (unchecked length)." Since this is a local pattern-based query, the trace is $\tau = \langle \ell \rangle$. Part (b) shows an inter-procedural use-after-free query: `isSource` matches the freed pointer and `isSink` matches its subsequent dereference, producing a full trace $\tau = \langle s, \ldots, \ell \rangle$.

```
1   // (a) CWE-120: unchecked length in memory-copy call
2   predicate isWriteFunc(Function f) {
3       f.getName().regexpMatch(
4           "(?i)^(memcpy|memmove|memset|strncpy|strncat)$") }
5   from FunctionCall fc, Function f, Expr n
6   where fc.getTarget() = f and isWriteFunc(f)
7       and countArg(fc, n)
8       and not isSizeBound(n) and not isSizeofLike(n)
9   select fc, "CWE-120: Buffer Overflow via "
10      + f.getName() + " (unchecked length)."
11  // (b) CWE-416: free-then-dereference data flow
12  module Cfg implements DataFlow::ConfigSig {
13      predicate isSource(DataFlow::Node s) {
14          exists(FunctionCall c, Expr a |
15              isDirectFreeCall(c, a) and s.asExpr() = a) }
16      predicate isSink(DataFlow::Node t) {
17          exists(Expr e | isDerefUse(e) and t.asExpr() = e) }
18  }
19  module DF = DataFlow::Global<Cfg>;
```

**Listing 2: Custom queries: (a) CWE-120 buffer overflow; (b) CWE-416 use-after-free.**

### 3.2 Fact Enrichment

Given a finding $f_i$, Sᴀɪʟᴏʀ enriches it into a *fact pack* of code-level hints by applying regex-based extractors to the function containing $\ell$ and the project build configuration:

- *Suspect calls*: function calls near $\ell$, extracted via identifier-followed-by-parenthesis pattern matching (e.g., `memcpy`, `malloc`, `free`).

**Table 1: Query suite.**

| Category | CWE / ID | # |
|---|---|---|
| *Standard (codeql/cpp-queries)* | | |
| Memory corruption | 120, 121, 125, 476, 787 | 6 |
| Integer overflow | 190 | 4 |
| UAF / double-free | 415, 416, 562 | 3 |
| *Custom rules* | | |
| Buffer overflow | 120, 125, 787, 823 | 6 |
| Null dereference | 476 | 1 |
| Uncontrolled recursion | 674 | 1 |
| Use-after-free | 416 (5 variants + realloc) | 6 |
| Stale pointer / type confusion | 416, 125 | 5 |
| Lifetime mismatch | 416 | 2 |
| **Total** | | **34** |

- *Pointer variables*: variables matching pointer or array declaration patterns.
- *Length variables*: identifiers whose names match size-related conventions (`len`, `size`, `count`, `capacity`).
- *Bounds hints*: bound relationships inferred from comparison expressions near $\ell$ (e.g., `ivlen >= 0`).
- *Build context*: include paths (`-I`) and preprocessor definitions (`-D`) extracted from the project's `compile_commands.json` (generated during the CodeQL database build), enabling Phase 2 to resolve transitive type dependencies.

For the running example, the suspect-call extractor identifies four function calls near $\ell$: `memcpy`, `bfd_zalloc`, `bfd_put_32`, and `bfd_set_section_alignment`. This directs the LLM toward `memcpy` as the dangerous copy operation and `bfd_zalloc` as the allocation site. The build context captures binutils' include paths (e.g., `-I bfd`), enabling the LLM to include the correct headers and resolve project-specific types when synthesizing the harness.

### 3.3 Vulnerability Specification Generation

Given a finding $f_i$ and its fact pack, this stage assembles a *vulnerability specification* by adding three elements:

(1) *Entrypoint selection*: Determines which project function serves as the symbolic execution entrypoint $e$. By default (LLM_INFER, used in all experiments), the orchestrator searches the call graph upward from the vulnerable function $f_v$ to find the nearest non-static caller as an initial $e$, then passes it to the LLM, which may override the choice during source exploration (§4.1) if it determines a different entry better reaches the vulnerability. The strategy is configurable (e.g., manual override or pure call-graph lookup).

**Table 2: Per-CWE assertion templates.**

| Vulnerability class | CWEs | Assertion template |
|---|---|---|
| Out-of-bounds read/write | 120, 121, 125, 787 | n <= min(len(dst), len(src)) |
| Use-after-free / double-free | 415, 416 | no use of p after free(p) |
| Stale pointer | 416 (variants) | p not dereferenced after realloc/free |
| Null dereference | 476 | p != NULL before *p |
| Integer overflow | 190 | arithmetic within type range |
| Out-of-range offset | 823 | offset within allocation bounds |
| Return of stack address | 562 | no return of stack-local address |
| Uncontrolled recursion | 674 | recursion depth bounded |

(2) *Assertion template*: The vulnerability specification includes a per-CWE assertion template (Table 2) that guides the LLM in

encoding the safety property during Phase 2 (§4.3). The CWE identifier is extracted from the query's `rule_id`; for well-known vulnerability classes, the template provides a precise safety condition (e.g., buffer bounds, use-after-free lifecycle), reducing LLM reasoning errors. For CWEs without a dedicated template, the LLM derives the property from the vulnerability description $d$ and the fact pack. For the running example, the CWE-120 template `n <= min(len(dst), len(src))` guides Phase 2 to bind n, dst, src to the `memcpy` arguments.

(3) *Filtering*: Non-actionable specifications are removed by matching file paths against skip patterns (test directories, benchmarks, examples, fuzz harnesses, generated files) and function names against exclusion patterns (main, test/benchmark prefixes, compiler-internal names). Project-specific CLI tool files are also excluded so that only core library code remains. For binutils, Fact Generation produces 19,140 findings; filtering reduces these to 1,260 active targets for Phase 2.

Figure 2 shows the resulting specification for the running example, assembling $\ell$, the description $d$, the suspect calls from the fact pack, the entrypoint selection LLM_INFER, and the assertion template into a single JSON document. Since each specification is self-contained, Phase 2 can process targets independently and in parallel, enabling SAILOR to scale to large codebases without requiring whole-program SE.

```
Vulnerability Specification

{ "rule_id": "local/cpp/cwe-120-overflow",
  "file": "bfd/elfxx-x86.c", "line": 2699,
  "message": "CWE-120: Buffer Overflow via memcpy (unchecked length).",
  "snippet": "memcpy(htab->..->contents, ..eh_frame_plt, ..->size);",
  "suspect_calls": ["memcpy","bfd_zalloc","bfd_put_32"],
  "entrypoint": "LLM_INFER",
  "assertion_template": "n <= min(len(dst), len(src))" }
```

**Figure 2: Phase 1 output for the running example.**

## 4 Phase 2: LLM Orchestrated Symbolic Execution

Phase 2 addresses the core synthesis problem: given a vulnerability specification (§3.3), construct a harness $H$ that a SE engine can use to determine whether the vulnerability condition at $\ell$ is satisfiable. SAILOR automates this with an *LLM-orchestrated* pipeline (Algorithm 1) that synthesizes a driver to set up the SE state, stubs to isolate the vulnerability path from irrelevant code, and assertions to encode the safety property, then iteratively refines them against compiler and symbolic execution feedback.

### 4.1 Driver Synthesis

The driver is the `main()` function that constructs an initial symbolic state from which KLEE explores paths toward $\ell$. During the first $T_{explore}$ turns, the LLM issues tool calls to retrieve function signatures, struct definitions, and type declarations from the project source (Algorithm 1, exploration phase). This *source exploration* phase provides the context needed for three key driver decisions:

(1) *Entry function $e$*: The LLM selects $e$ based on the entrypoint strategy from Phase 1 (§3.3) and retrieves its signature to determine parameter types and allocate arguments.

**(a) Prompt**

```
SPEC:
  l: elfxx-x86.c:2699
  e: _bfd_x86_elf_late_size_sections
  d: CWE-120: unchecked memcpy length
  guards: plt_eh_frame==NULL; contents==NULL
  assert: n <= min(len(dst), len(src))
DRIVER RULES:
  Entry: call e with allocated parameters
  Symbolic: overapproximate; make fields symbolic
    GOOD: klee_make_symbolic(&ctx, sizeof(ctx), "ctx"); BAD: ctx->state = 7;// hardcoded
  Guards: for each guard c, add klee_assume(!c)
    GOOD: klee_assume(ptr != NULL);
  Pointers: real allocs, never NULL
    GOOD: p = calloc(1, sizeof(*p));
  Buffers: alloc + make contents symbolic
    GOOD: buf = malloc(N); klee_make_symbolic(buf, N, "buf");
```

**(b) Generated driver**

```
1  int main() {
2      bfd *out = calloc(1, sizeof(bfd));              // (1) alloc
3      struct bfd_link_info *info = calloc(1, sizeof(struct bfd_link_info));
4      global_htab = calloc(1, sizeof(struct elf_x86_link_hash_table));
5      global_htab->plt_eh_frame = calloc(1, sizeof(asection)); // (2) guard
6      unsigned char *dst = malloc(16);                // (3) sym buf
7      klee_make_symbolic(dst, 16, "dst_bytes");
8      global_htab->plt_eh_frame->contents = dst;      // (4) guard
9      unsigned char *src = malloc(512);
10     klee_make_symbolic(src, 512, "src_bytes");
11     global_htab->plt_eh_frame_plt = src;
12     size_t sz;                                      // (5) sym scalar
13     klee_make_symbolic(&sz, sizeof(sz), "copy_size");
14     global_htab->plt_eh_frame->size = sz;
15     _bfd_x86_elf_late_size_sections(out, info);     // (6) call e
16     return 0; }
```

**Figure 3: Driver synthesis for the running example. In (a), bold = template rules, <span style="color:purple">purple</span> = instantiated values from the specification; (b) shows the LLM-generated driver.**

---

**Algorithm 1:** LLM-orchestrated harness synthesis.

**Input:** Vulnerability specification $spec_t$, project source $P$
**Output:** Concrete test inputs or *inconclusive*
**Data:** $T_{explore}$: exploration budget (default 8), $T_{author}$: authoring deadline (default 12), $T_{max}$: total turn budget (default 60), $R_{max}$: refinement limit after site reached

```
1   t ← 0;  refine_count ← 0;
2   while t < Tmax do
3       action ← LLM(plan, history);
4       if t < Texplore then
5           // Source exploration
            Execute source search / read / extract;
6       else if t < Tauthor then
7           // Harness authoring
            Write or modify driver, code slice, stubs, assertions;
8       else
9           // Harness refinement
            result ← COMPILEDIAGNOSE(H, P);
10          if result = diagnostic then
11              Feed diagnostic to LLM;
12              t ← t + 1;  continue;
            // SE feedback
13          (outcome, diag) ← EXECDIAGNOSE(bitcode);
14          if outcome = bug_triggered then return concrete test inputs;
15          if outcome = site_reached then
16              refine_count ← refine_count + 1;
17              if refine_count > Rmax then return likely false positive;
18          Feed diag back to LLM;
19      t ← t + 1;
20  return inconclusive;
```

These decisions are encoded as prompt rules (Figure 3a), the vulnerability specification from Phase 1 is included verbatim, followed by driver construction rules with examples distinguishing correct patterns (symbolic + `klee_assume`) from common mistakes (hardcoded values). Figure 3b shows the resulting LLM-generated driver for the running example, where the LLM selects the three fields needed to reach the memcpy: `plt_eh_frame` and `contents` as concrete pointers satisfying the guards, and `size` as a symbolic scalar.

## 4.2 Stub Synthesis

The LLM constructs a self-contained *code slice*: A C file containing only the code along the call chain from $e$ to $\ell$, with all external dependencies replaced by stubs. The same source exploration that identified guard conditions for the driver (§4.1) also identifies *off-path* branches that do not contain the target statement and callees not needed to reach $\ell$. The LLM synthesizes stubs at four granularities:

(1) *Function-level:* Off-path callees are replaced with stubs matching the original signature. If a stub's return value influences the path to $\ell$ (e.g., controls a branch condition), it returns a symbolic value via `klee_make_symbolic`; otherwise it returns a hardcoded default. CWE-specific exceptions: for use-after-free targets, the `free()` stub must call the real `free()` so the free-then-use sequence remains intact.

(2) *Branch-level:* Off-path `if` blocks are replaced with `if(0)`, and off-path `switch` cases with `break`.

(3) *Loop-level:* Loops enclosing the target statement are converted to single-pass `if` conditionals, bounding KLEE's path count.

(4) *Type-level:* Project structs are redefined to include only the fields accessed by the sliced code, avoiding transitive type dependencies.

Figure 4 shows the prompt rules and the resulting code slice for the running example. In the original function, a 100+ line if-block and several callees are off-path, while the memcpy is the on-path target. In the generated slice (Figure 4b), the 40+ field struct is reduced to 3 fields (type-level), `bfd_get_32` returns a symbolic value (function-level), the off-path block becomes `if(0)` (branch-level), and the enclosing loop becomes `if(1)` (loop-level).

These decisions are encoded as prompt rules (Figure 3a), the vulnerability specification from Phase 1 is included verbatim, followed by driver construction rules with examples distinguishing correct patterns (symbolic + `klee_assume`) from common mistakes (hardcoded values). Figure 3b shows the resulting LLM-generated driver for the running example, where the LLM selects the three fields needed to reach the memcpy: `plt_eh_frame` and `contents` as concrete pointers satisfying the guards, and `size` as a symbolic scalar.

(2) *Symbolic inputs:* The LLM retrieves the type definitions of $e$'s parameters and the struct fields referenced along the path from $e$ to $\ell$. Using the vulnerability specification (§3.3), it partitions fields into three categories: (a) *Symbolic scalars:* Fields that appear in the vulnerability condition are declared as symbolic via `klee_make_symbolic`; (b) *Concrete pointers:* Pointer fields are allocated via `calloc/malloc` so that KLEE can dereference them without triggering spurious null-pointer errors; and (c) *Symbolic buffers:* Heap regions whose *contents* are declared symbolic while the pointer itself remains concrete.

(3) *Guard conditions:* The LLM identifies conditional statements along the call chain from $e$ to $f_v$ that cause early exit before reaching $\ell$ (e.g., `if (p == NULL) return`). For each guard with condition $c$, the driver encodes `klee_assume(¬c)` to bypass the early-exit path.

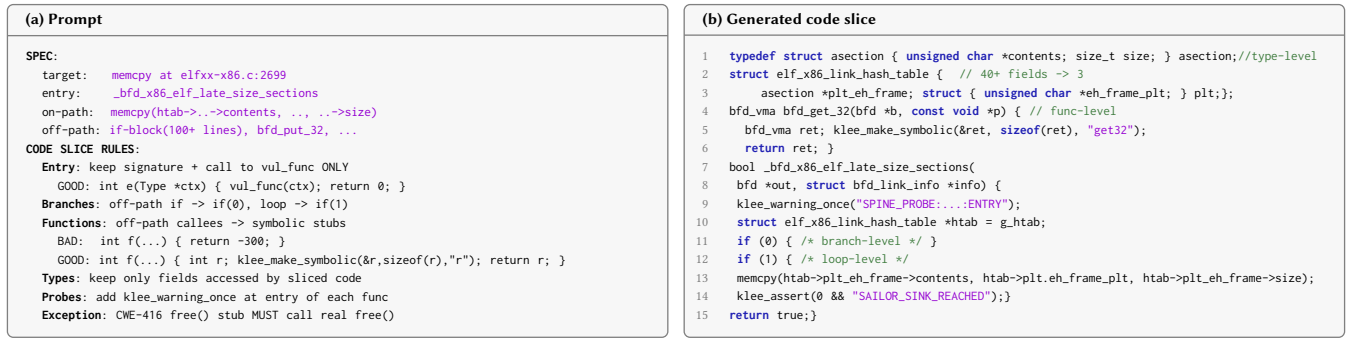

**Figure 4: Stub synthesis for the running example. In (a), bold = template rules, purple = instantiated values; (b) shows the LLM-generated code slice.**

### 4.3 Assertion Instantiation

The assertion template from the vulnerability specification (§3.3) specifies *what* safety property to check; this stage determines *how* to encode the safety property in $H$ so that KLEE can verify it. Two complementary mechanisms are used:

**Reachability assertion.** A `klee_assert(0)` is placed immediately after the vulnerable statement in the code slice. KLEE fires this assertion only if the statement executes without a memory error, confirming that $\ell$ is reachable from the driver.

**Vulnerability condition.** The LLM reads the assertion template and adds `klee_assume` constraints to the driver that violate the safety property, relying on KLEE's native error detection (out-of-bounds access, use-after-free, null dereference) to flag the violation at $\ell$. Figure 5 shows the prompt rules and the resulting instantiation patterns for three vulnerability families. For the running example, the template `n <= min(len(dst), len(src))` identifies three operands. The initial driver (Figure 3b) already declares `sz` as symbolic and allocates `dst` (16 bytes) and `src` (512 bytes). To complete $H$, the LLM adds `klee_assume(sz > 16)` to violate the bound and `klee_assume(sz <= 512)` to keep the access within the source buffer. KLEE then detects the out-of-bounds write at the `memcpy` call.

### 4.4 Harness Refinement

Once the LLM produces an initial $H$, the orchestrator enters a compile-execute-refine loop (Algorithm 1, **else** branch) that feeds diagnostics back to LLM so it can fix the driver, stubs, or assertions.

*Compilation feedback.* $H$ is compiled to LLVM bitcode (`clang -O0 -g`) and linked into a single module (`llvm-link`). If compilation fails, the orchestrator pattern-matches the error into one of four classes: *incomplete type*, *conflicting prototype*, *redefinition*, or *other* and augments the diagnostic with a suggested fix (e.g., looking up the missing struct field from project headers, or grepping for the real prototype). The LLM receives the augmented error and fixes $H$ accordingly.

*Symbolic execution feedback.* Once compilation succeeds, KLEE explores the bitcode with dual-strategy search (random-path + depth-first), a 300 s per-run timeout, and a depth limit of 1,000.

The orchestrator classifies each run into one of three outcomes: (i) *not reached*: $\ell$ was not executed; the orchestrator reports which functions in the call chain were entered (via `klee_warning_once` coverage probes) and which were not, enabling the LLM to fix the driver or stubs; (ii) *site reached*: `klee_assert(0)` fires, confirming reachability but no memory safety violation; the LLM may tighten driver constraints; (iii) *bug triggered*: KLEE detects a memory safety violation at $\ell$ and produces concrete witness inputs (`.ktest` files).

*Termination.* The loop terminates when a bug is triggered (producing concrete test inputs for Phase 3), the iteration budget is exhausted (default 60 turns, marking the specification as *inconclusive*), or the site is reached but no bug is triggered after 15 additional refinement turns (classifying it as a likely false positive). Figure 6 shows both types of refinement feedback for the running example. When the bug is triggered, Phase 2 produces the output artifacts (Figure 7).

## 5 Phase 3: Concrete Validation

For each specification where KLEE detected a memory error, this phase validates the output by replaying the concrete witness inputs against the *unmodified project source*. This helps to eliminate false positives that may be due to unrealistic harness generation by LLM. First, the symbolic driver is transformed into a concrete replay driver: each `klee_make_symbolic` call is replaced with a `memcpy` of the witness bytes from the `.ktest` file, and all remaining KLEE-specific calls are removed. Then the unmodified project source is compiled with AddressSanitizer (`-fsanitize=address`), producing an instrumented static archive (`.a`). The replay driver is linked against this `.a` and executed. A crash is classified as *confirmed* only if the ASan stack trace reports a memory safety violation inside the project's source code.

Each pipeline phase characterizes the vulnerability at a different level of abstraction: CodeQL reports CWE-level pattern (e.g., CWE-120), KLEE reports memory error (`.ptr.err` or `.free.err`), and ASan reports the concrete violation (e.g., `heap-buffer-overflow`). These classifications are complementary; the ASan verdict serves as the ground truth because it is produced by executing the unmodified code. For each confirmed bug, this phase outputs a concrete replay driver, ASan stack trace, and a final verdict report. Figure 8 shows the Phase 3 output for the running example: ASan confirms a `heap-buffer-overflow` in `_bfd_x86_elf_late_size_sections`, reading 8 bytes past a 4-byte allocation. The `verified_bug.json` is the final pipeline output, assembling the ASan verdict with the symbolic witness inputs. Together with the Phase 1 specification (Figure 2) and Phase 2 KLEE artifacts (Figure 7), these artifacts provide a reproducible evidence package for the detected vulnerability.

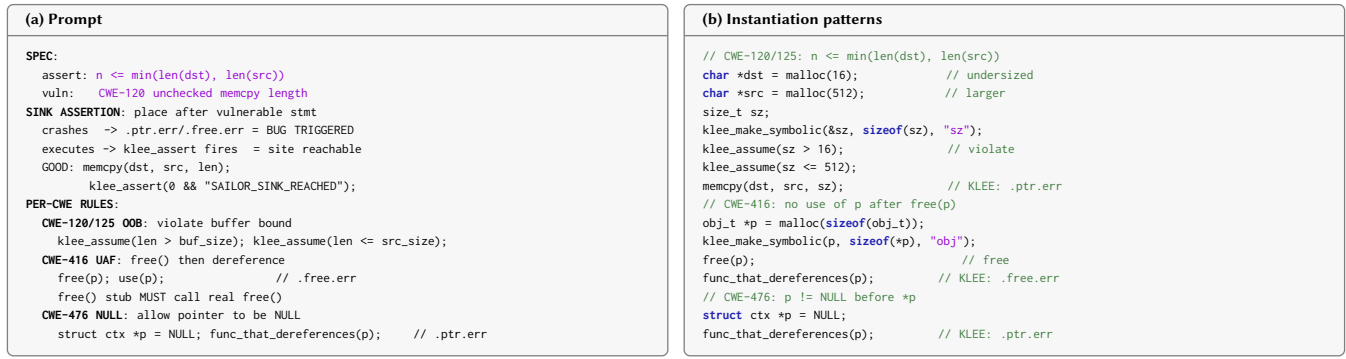

**Figure 5: Assertion instantiation. In (a), bold = template rules, purple = instantiated values; (b) shows per-CWE encoding patterns.**

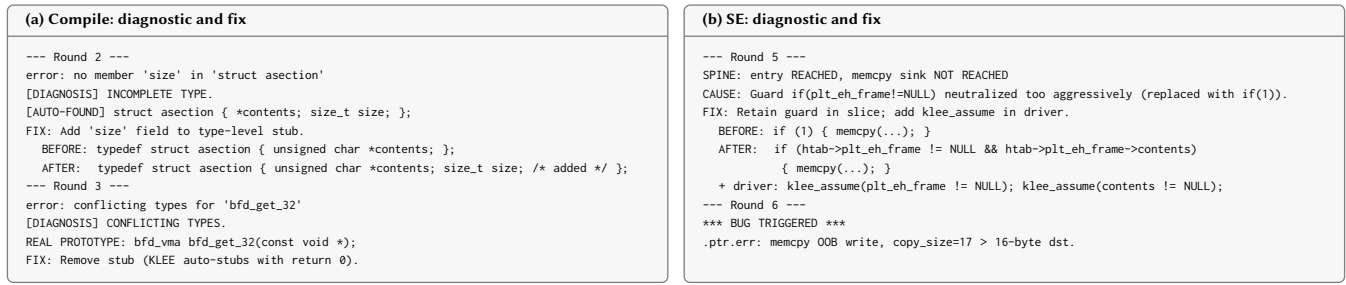

**Figure 6: Harness refinement for the running example: compilation feedback (a) fixes type-level stubs; SE feedback (b) restores over-neutralized guards, leading to bug trigger.**

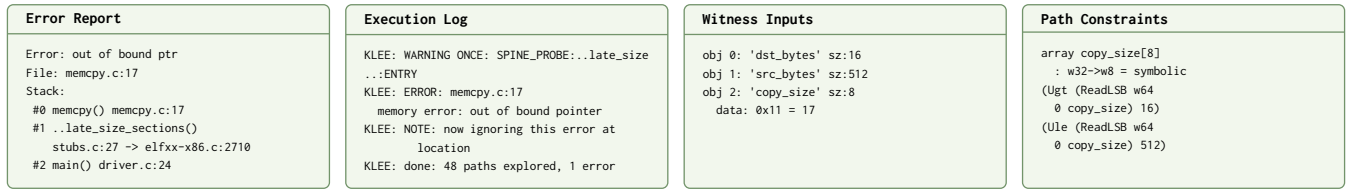

**Figure 7: Phase 2 output for the running example.**

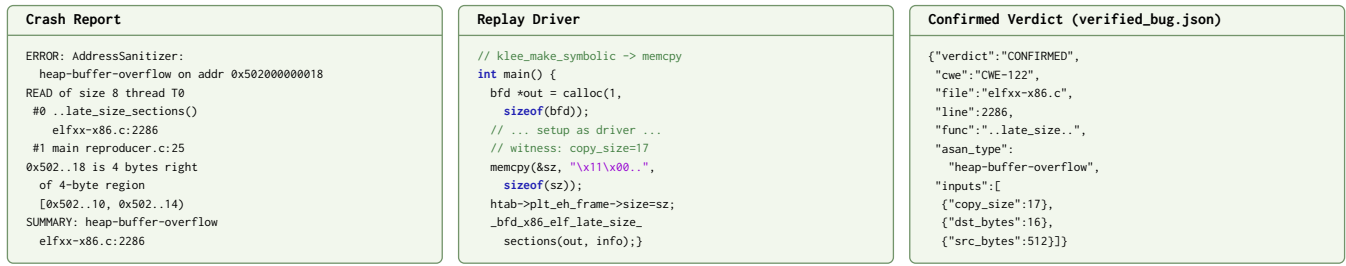

**Figure 8: Phase 3 output for the running example.**

## 6 Evaluation

We evaluate SAILOR on 10 real-world, open-source C/C++ projects spanning image processing (libpng, libtiff), parsing (libxml2), networking (curl), security (OpenSSL, SELinux), multimedia (FFmpeg), document rendering (mupdf), binary analysis (binutils), and databases (SQLite) at recent commits, totaling 6.8 M LOC across 14,634 source files (Table 4). We address three research questions:

- **RQ1 (Effectiveness):** How many unique, confirmed vulnerabilities does SAILOR discover?
- **RQ2 (Comparison):** How does SAILOR compare against alternative vulnerability discovery approaches?
- **RQ3 (Ablation):** How much does each component of SAILOR contribute to the end-to-end results?

### 6.1 Experimental Setup

*Baselines.* We compare SAILOR against five baselines and one ablation (Table 3):

**B1: SE with human-written harnesses.** We collect existing OSS-Fuzz harnesses and convert them to KLEE-compatible drivers by replacing the fuzz inputs with symbolic variables. For projects without OSS-Fuzz coverage (libpng, SQLite, mupdf), we manually write harnesses targeting core API entrypoints. Each harness is compiled and linked against the full project source. In total, there are 86 harnesses (76 from OSS-Fuzz, 10 manually written).

**B2: SE with LLM-generated harnesses.** We extract API prototypes from public headers and prompt the LLM with the prototypes and target CWE types to generate KLEE harnesses (driver, stubs, and assertions). The returned C files are compiled to LLVM bitcode, linked with the project bitcode, and executed under KLEE with no further LLM interaction.

**B3: LLM vulnerability detection.** Each source file is individually sent to the LLM along with the target CWE types. The LLM is prompted: "Identify memory-safety vulnerabilities and provide the entry function, argument values, and trigger condition that crash this code."

**B4: SA-guided LLM vulnerability detection.** CodeQL findings are grouped by source file. For each file, the LLM is prompted with all raw findings, the source code, and the target CWE types: "Triage each finding as true/false positive; for true positives, provide the entry function and crashing inputs." Unlike SAILOR, findings are not converted into vulnerability specifications.

**B5: Agentic LLM vulnerability detection.** An autonomous agent with full codebase access (file read, grep, shell, compilation) and unlimited turns is prompted with the target CWE types: "Find memory-safety vulnerabilities and provide entry functions and crashing inputs." The agent decides its own exploration strategy.

### Table 3: Baseline and ablation configurations.

| ID | Approach | SA | LLM | SE | LLM context | LLM budget |
|---|---|---|---|---|---|---|
| B1 | SE with human-written harnesses | ✗ | ✗ | ✓ | – | – |
| B2 | SE with LLM-generated harnesses | ✗ | GPT-5 | ✓ | API headers | 600 s/harness |
| B3 | LLM vulnerability detection | ✗ | GPT-5 | ✗ | source files | 600 s/file |
| B4 | SA-guided LLM vulnerability detection | ✗ | GPT-5 | ✗ | SA findings | 600 s/file |
| B5 | Agentic LLM vulnerability detection | ✗ | Opus 4.6 | ✗ | full project | ∞/project |
| A1 | SAILOR without SA | ✓ | GPT-5 | ✓ | source files | 600 s/file |
| SAILOR | Full pipeline | ✓ | GPT-5 | ✓ | SA specs | 600 s/spec |

SE uses KLEE with 300 s timeout per run.
A1/SAILOR: 60 iterative turns with build+KLEE feedback.

### Table 4: Benchmark projects and SAILOR results. #SA: specs generated from CodeQL findings; #SE: vulnerabilities triggered by SE; #Conf.: confirmed crashes against real .a; #Uniq.: deduplicated by (file, function, line); #Fuzz: reproduced via fuzzing; Tok.: LLM tokens (millions).

| Project | Commit | LOC | #SA | #SE | #Conf. | #Uniq. | #Fuzz | Tok. (M) |
|---|---|---|---|---|---|---|---|---|
| libpng | 747dd02 | 63K | 609 | 44 | 29 | 21 | 0 | 24.0 |
| libtiff | f324415 | 95K | 1,491 | 71 | 21 | 14 | 15 | 39.1 |
| libxml2 | e334a9d | 149K | 1,987 | 377 | 16 | 11 | 10 | 320.7 |
| curl | 2eebc58 | 174K | 2,040 | 3 | 0 | 0 | 0 | 75.1 |
| SELinux | ca10fc4 | 190K | 12,498 | 285 | 62 | 62 | 41 | 161.6 |
| mupdf | 21fb0a2b | 1.25M | 6,775 | 187 | 141 | 141 | 80 | 122.3 |
| FFmpeg | f46e5144 | 1.3M | 29,697 | 109 | 78 | 78 | 54 | 581.2 |
| binutils | b2bc71a | 1.84M | 19,140 | 257 | 74 | 52 | 51 | 186.0 |
| OpenSSL | 67b5686b | 710K | 3,977 | 0 | 0 | 0 | 0 | 237.7 |
| SQLite | 0f88d958 | 1.05M | 9,171 | 12 | 0 | 0 | 0 | 540.1 |
| **Total** | | **6.8M** | **87,385** | **1,345** | **421** | **379** | **251** | **2,288** |

**A1: SAILOR without SA (ablation).** Runs the full SAILOR loop but without static analysis: no CodeQL-derived target locations, rule details, call chains, or bounds hints. The LLM sees only raw source code, CWE types, and its own scan results.

SE-based detection approaches (B1, B2, A1, SAILOR) output concrete witness inputs (.ktest) and LLM-based detection approaches (B3–B5) output crashing inputs directly. All outputs are validated against the project's unmodified .a (§5); only crashes whose stack trace enters the .a are counted.

*Infrastructure.* All experiments run on two identical servers (Intel Xeon Platinum 8462Y+, 128 threads, 251 GB RAM, Ubuntu 22.04) inside Docker containers. B5 uses `claude-opus-4-6` as the strongest available agentic LLM; all other LLM-based approaches use `gpt-5-0806` for consistency.

### 6.2 RQ1: Effectiveness

Table 4 summarizes SAILOR's results across all 10 projects. SAILOR produced **421 confirmed crashes**, corresponding to **379 unique vulnerabilities** after deduplication by (file, function, line). Of the 421 confirmed, 251 (60%) were reproduced by seeding OSS-Fuzz-compatible libFuzzer with the concrete witness inputs from KLEE. The remaining 170 were not fuzz-reproduced because the triggering condition requires precise multi-field struct initialization that random mutation is unlikely to reconstruct (e.g., libpng's 29 vulnerabilities require specific chunk-type and bit-depth combinations). Of the 87,385 static analysis findings, 1,345 were detected as vulnerabilities by symbolic execution. By ASan error type, the confirmed crashes comprise heap-buffer-overflow (288, 68%), use-after-free (56, 13%), segmentation fault (53, 13%), stack-buffer-overflow (15, 4%), double-free (4, <1%), and other (5, 1%). Three projects produced 0 confirmed vulnerabilities, each for a distinct reason: *curl* (SE detected 3, 0 confirmed), the library requires multi-step session setup (e.g., `curl_easy_init` → `curl_easy_setopt` → callback registration) that the LLM could not complete within the 60-turn budget. *OpenSSL* (0/3,977 specs compiled for SE): all harnesses failed to compile due to OpenSSL's complex internal type hierarchy and multi-step API initialization (e.g., `EVP_CIPHER_CTX_new` → `EVP_-EncryptInit_ex` with engine and key setup) that the LLM could not resolve within the turn budget. *SQLite* (12 detected by SE, 0 confirmed): KLEE triggered errors (e.g., `btree.c:4074`) but internal functions like B-tree operations require a valid database state (open handle, schema, page cache) that raw KLEE byte values cannot reconstruct during ASan replay.

**Table 5: Comparison of all approaches. Each cell shows *confirmed / detected*. B1–B2 parenthetical = harnesses tried (\* = manually written, no OSS-Fuzz). All confirmed vulnerabilities crash inside the unmodified project `.a`.**

|         | B1       | B2      | B3     | B4      | B5      | A1     | Sailor   |
|---------|----------|---------|--------|---------|---------|--------|----------|
| libxml2 | 0/11 (12) | 0/2 (28) | 0/345  | 0/49    | 0/5     | 20/47  | 11/377   |
| libtiff | 0/2 (2)  | 0/6 (19) | 5/72   | 0/64    | 0/7     | 11/53  | 14/71    |
| libpng  | 0/0(4\*) | 0/0(20) | 0/162  | 0/2     | 0/29    | 0/40   | 21/44    |
| binutils | 0/1 (15) | 0/0(30) | 0/60   | 0/36    | 0/63    | 0/0    | 52/257   |
| curl    | 0/5 (6)  | 0/0 (25) | 0/138  | 0/28    | 0/7     | 0/9    | 0/3      |
| OpenSSL | 0/0(31)  | 0/0(27) | 0/818  | 0/194   | 5/157   | 0/14   | 0/0      |
| FFmpeg  | 0/0(6)   | 0/0(21) | 0/57   | 0/680   | 0/0     | 0/3    | 78/109   |
| SELinux | 0/0(4)   | 0/0(26) | 0/30   | 2/81    | 6/82    | 0/19   | 62/285   |
| SQLite  | 0/0(3\*) | 0/32(36) | 0/99   | 0/10    | 1/43    | 0/72   | 0/12     |
| mupdf   | 0/0(3\*) | 0/4(49) | 0/0    | 0/219   | 0/37    | 0/19   | 141/187  |
| **Total** | **0/19** | **0/86** | **5/1781** | **2/1363** | **12/430** | **31/276** | **379/1345** |

The Sailor pipeline processed 87,385 vulnerability specs, consuming 2.29 B LLM tokens across all 10 projects (Table 4). Per-spec token cost averages 26 K tokens/spec. Per-bug cost ranges from 0.9 M tokens (mupdf) to 29.2 M tokens (libxml2), and does not correlate with project size, but depends on harness complexity. Projects with well-isolated functions (mupdf, libpng) require fewer iterations than those with deep API dependencies (libxml2, FFmpeg). Projects with zero confirmed vulnerabilities (curl, OpenSSL, SQLite) consume 852.9 M tokens (37% of total) because the LLM spends its full turn budget attempting harnesses that ultimately fail to compile or produce non-reproducible witness values. Each spec runs as an independent session, averaging 2.7 minutes wall-clock time with 128 parallel workers.

---

**RQ1:** Sailor discovers 379 unique vulnerabilities (421 confirmed crashes) across 7 of 10 projects. The pipeline scales to projects up to 1.84 M LOC (binutils: 52 vulnerabilities) but struggles with complex API initialization (curl, OpenSSL: 0 vulnerabilities). Cost depends on harness complexity, not project size (0.9 M tokens/bug for mupdf vs. 29.2 M for libxml2).

---

### 6.3 RQ2: Comparison with Baselines

Table 5 compares Sailor against all baselines on confirmed unique vulnerabilities.

*B1:* For B1, across 86 harnesses, 0 confirmed vulnerabilities are detected. These harnesses target whole-project or API-level entry-points, producing state spaces too large for symbolic execution; 53 of 86 (62%) fail to compile.

*B2:* Of 281 generated harnesses in B2, 228 (81%) fail to compile due to LLM hallucinations (incorrect function signatures, missing types, and fabricated APIs) that compilation feedback would have caught. The 53 that compile trigger 86 KLEE errors, but all crashes occur in harness setup code.

*B3:* B3 reports 1,781 potential vulnerabilities, of which only **5 unique vulnerabilities** (all in libtiff) survive `.a` validation: 1 null dereference in `TIFFUnRegisterCODEC` and 3 negative-size `memcpy/memset` in `tif_unix.c`, all shallow functions with a direct argument-to-memory path. Deeper vulnerabilities require path constraints the LLM cannot reason about without SE.

*B4:* B4 produces 1,363 reported vulnerabilities, but only **2** (both in SELinux) survive `.a` validation: a use-after-free in `mls_semantic_-level_cpy` and a stack overflow in `sepol_node_get_addr_bytes`. Raw SA findings (up to 19K for binutils) exceed the LLM's context capacity; Sailor's spec generation (§3.3) addresses this by converting each finding into vulnerability specification with vulnerability context, entrypoints, and assertion templates.

*B5:* B5 confirms **12 unique vulnerabilities** across 3 of 10 projects (OpenSSL: 5, SELinux: 6, SQLite: 1), with 11 *exclusive* to B5. These vulnerabilities reside in API-boundary functions (e.g., negative `flen` in RSA padding, `szDb` > `szBuf` in SQLite memdb) where the agent crafts semantically meaningful inputs that SE cannot derive. However, B5 confirms 0 vulnerabilities in 7 projects. Of 425 crashing inputs across 10 projects, only 105 (25%) trigger a crash. Of those 105 crashes, 51% are duplicates, the agent repeatedly targets the same location (e.g., all 11 SQLite crashes hit `memdb.c:266`). After deduplication, 51 unique crash locations remain, of which only 12 survive `.a` validation. The remaining inputs do not reach the suspected vulnerable point, for example, in libpng the agent targets `png_set_hIST` but the input does not pass through the required `png_create_info_struct` initialization.

*A1:* A1 confirms **31 unique vulnerabilities** (libxml2: 20, libtiff: 11) with zero overlap with Sailor's 379 on (file, line): the LLM targets parser functions (e.g., `tif_ojpeg`, `HTMLparser`) while CodeQL targets buffer operations (e.g., `tif_swab`, `encoding.c`). The ablation impact is analyzed in RQ3. The union of all approaches yields 427 unique vulnerabilities; Sailor alone covers 89% (379/427).

---

**RQ2:** No single baseline exceeds 12 unique vulnerabilities whereas Sailor finds 379. SE-based baselines (B1–B2) fail due to harness quality; LLM-only baselines (B3–B5) fail due to high false positive rates (88–99%).

---

### 6.4 RQ3: Ablation Study

To understand which Sailor components contribute most, we analyze each phase's contribution and the ablation results.

*SA targeting.* Of 87,385 SA findings across all projects, SE detected the target vulnerable line in 1,345 cases, with conversion rates ranging from 1.3% (binutils) to 7.2% (libpng). The A1 (Table 5) confirms that removing SA targeting and relying on LLM source scanning reduces confirmed vulnerabilities from 379 to 31, a 12.2× drop.

**Table 6: Harness quality across all 10 projects.**

|  | Total | Compiled | Detected | Confirmed |
|---|---|---|---|---|
| B1 | 86 | 38% | 22% | 0% |
| B2 | 281 | 19% | 31% | 0% |
| Sailor | 87,385 | 44% | 1.5% | 0.5% |

*Iterative refinement.* The compile-execute-refine loop is critical. Table 6 compares harness quality across all 10 projects. B1's human-written harnesses compile at 38% but confirm nothing, they target API-level entrypoints too broad for SE. B2's one-shot harnesses compile at only 19% due to LLM hallucinations, and detected vulnerabilities all crash in harness code. Sailor's iterative loop compiles at 44% and confirms 421 vulnerabilities, averaging 8.4 KLEE runs per specification.

*SE necessity.* SA alone produces 87,385 findings with a false positive rate of 99.6% (379 confirmed). LLM detections do not close this gap: B4 (SA+LLM triage) reports 1,363 with only 2 confirmed (99.9% FP), B3 (LLM-only) reports 1,781 with 5 confirmed (99.7% FP), and B5 (agentic) triggers 105 crashes with 12 confirmed (88.6% FP). SE triggers vulnerabilities in 1,345 specifications, of which 421 are concretely confirmed. Of the 924 unconfirmed, 19 are identified as crashes in LLM-generated stubs; the remaining 905 are inconclusive: KLEE witness values that do not reproduce the crash under ASan replay against the real .a.

*Concrete validation.* Concrete validation is conservative: it rejects vulnerabilities where witness inputs do not trigger the crash through the real .a's initialization path. For example, SQLite's memdb.c:266 (found by B5 via sqlite3_deserialize with szDb > szBuf) is detected by SE in Sailor but the KLEE-generated byte values cannot reconstruct the required database state for replay.

> **RQ3:** Each component is necessary: removing SA reduces vulnerabilities 12.2× (A1), removing iterative feedback reduces them to 0 (B2), and removing SE reduces them to ≤12 (B3–B5). Concrete validation filters 924 unconfirmed SE findings, yielding 379 unique confirmed vulnerabilities.

## 6.5 Discussion

*vulnerabilities SAILOR missed.* Across B3 and B5, 15 unique vulnerabilities were found that Sailor missed. We classify them by root cause:

- *SA gap* (4/15 = 27%): CodeQL did not flag the vulnerable location. In OpenSSL's bf_readbuff.c, the root cause is an integer overflow at line 92 where INT_MAX+1 wraps the buffer size to zero; CodeQL's CWE-125 rule flagged only the downstream buffer read at line 125, so Sailor targeted the wrong line. Three additional SELinux vulnerabilities found by B5 were not covered by the current rule suite. All SA gaps are addressable by expanding or refining rules.

- *KLEE limitation* (11/15 = 73%): CodeQL correctly flagged the locations, but KLEE could not produce crashing concrete values. Causes include negative-value edge cases (OpenSSL RSA flen, 3 vulnerabilities), imprecise .ktest replay (SQLite memdb, 1 bug), and complex API state initialization that the LLM could not resolve within the turn budget (OpenSSL EVP, SELinux policydb, libtiff wrappers; 7 vulnerabilities).

*LLM influence.* We evaluated Sailor with DeepSeek-V3.2 on libtiff under the same configuration as GPT-5. DeepSeek-V3.2 produces 490 SE-detected findings versus 71 with GPT-5, but confirms only 12 unique vulnerabilities versus 14. The LLM's code reasoning capability influences harness quality, but the pipeline's iterative feedback loop and concrete validation compensate: even with a weaker LLM, Sailor still confirms 86% (12/14) of the vulnerabilities found by the stronger model.

*Threats to validity. Internal:* Deduplication by (file, function, line) may over- or undercount root causes. *External:* Results are specific to 10 C/C++ projects, memory safety CWEs, and the GPT-5/Opus 4.6 model snapshots. *Data contamination:* The LLM may have seen the target projects during training, but concrete validation against the unmodified .a is LLM-independent, and the DeepSeek-V3 experiment confirms similar results with a different model.

## 7 Related Work

*Harness generation for SE.* STASE [29] uses SA to identify targets for SE in UEFI firmware but requires harnesses written from fixed templates with manually specified entrypoints and environment models. Sys [5] uses under-constrained SE guided by programmer-written checkers. Busse et al. [6] feed SA error traces to SE but report that the traces are rarely useful. AutoBug [15] replaces the SE engine with LLM-powered symbolic reasoning but has not been demonstrated on large real-world codebases. A recent poster [35] explores LLM-enhanced KLEE without a full pipeline or evaluation. No prior work uses LLMs to automatically construct SE harnesses.

*Fuzz driver generation.* In contrast to SE harnesses, automated fuzz driver generation has received considerable attention. Traditional approaches extract API usage from consumer code [2, 13], unit tests [14], dynamic traces [16, 36], or Android/Windows interfaces [16, 22, 32]; Hopper [9] uses a DSL to avoid needing consumer code. LLM-based approaches [12, 23, 24, 31, 33, 34, 37] generate drivers with increasing sophistication from zero-shot generation [31] to coverage-guided prompt mutation [24] and agentic pipelines [34] but all target coverage-guided fuzzing, not symbolic execution.

*LLM-assisted SA.* LLMs have been combined with SA for vulnerability detection: IRIS [21] infers taint specifications for CodeQL; LLift [19] improves kernel bug detector precision; others triage SA alerts [17]. These use LLMs for *detection* without execution-based confirmation. Sailor uses SA alerts as input to LLM harness *construction*, with SE and concrete validation providing the confirmation. No existing system combines SA-informed targeting, LLM-driven harness synthesis, and concrete validation.

## 8 Conclusion

We presented Sailor, the first pipeline to fully automate symbolic execution harness construction for large C/C++ codebases. By combining static analysis informed targeting, LLM-orchestrated harness synthesis, and concrete validation, Sailor discovers 379 unique previously unknown vulnerabilities across 10 projects, confirming that the three techniques are more effective together than any one alone. While our evaluation uses CodeQL, GPT-5, and KLEE, the pipeline design separates static analysis targeting, LLM orchestration, and symbolic execution into independent phases, facilitating adaptation to other tools and LLMs.

## 9 Data Availability

Sailor implementation, baselines, and reproduction instructions are publicly available at https://doi.org/10.5281/zenodo.19344146. Crash reproducers and vulnerability details are withheld because the discovered vulnerabilities are being disclosed to the affected maintainers, and will be released once disclosure is complete.

## Acknowledgements

Per ACM policy, we disclose that GPT-5, DeepSeek-V3.2, and Claude Opus 4.6 are used as components of the Sailor pipeline and baselines (§6).